\documentclass[12pt]{article}
\usepackage{amsmath,amsthm,amssymb,amscd,epsfig}
\newtheorem{theorem}{Theorem}[section]

\newtheorem{lemma}[theorem]{Lemma}

\def\dim {{\sl Proof.}\phantom{X}}

\def\hk{{\cal H}_N(\kappa)}

\font\twelvebb=msbm10 scaled 1100 \font\fivebb=msbm5 \font\sevenbb=msbm7
\newfam\bbfam  \scriptscriptfont\bbfam=\fivebb
\textfont\bbfam=\twelvebb \scriptfont\bbfam=\sevenbb
\def\bb{\fam\bbfam\twelvebb}
\def\N{{\bb N}}
\def\Z{{\bb Z}}
\def\R{{\bb R}}

\def\T{{\bb T}}

\def\tor{{{\T^{2d}}}}

\def\hh{{\cal H}_{\hbar}(\kappa)}

\def\e#1{e^{\displaystyle #1}}

\def\opwk#1{{Op^W_\kappa(#1)}}
\def\opawk#1{{Op^{AW}_{\kappa}(#1)}}

\def\norm#1{{\left\|{#1}\right\|}}

\begin{document}

\title{Controlling strong scarring for \\ quantized ergodic toral automorphisms}
\author{Francesco Bonechi\\
INFN, Sezione di Firenze \\Dipartimento di Fisica, Universit\`a di Firenze \\
Via Sansone 1, 50019 S.Fiorentino (Firenze), Italy\\ e-mail: bonechi@fi.infn.it\\ and\\
Stephan De Bi\`evre\\ UFR de Math\'ematiques et UMR AGAT\\Universit\'e des
Sciences et Technologies de Lille\\ 59655 Villeneuve d'Ascq Cedex
France\\e-mail: debievre@gat.univ-lille1.fr }
\date{\today}

\maketitle
\begin{abstract} We show that in the semi-classical limit the eigenfunctions
of quantized ergodic symplectic toral automorphisms can not concentrate in
measure on a finite number of closed orbits of the dynamics. More generally,
we show that, if the pure point component of the limit measure has support  on
a finite number of such orbits, then the  mass of this component must be
smaller than two thirds of the total mass. The proofs use only the algebraic
({\em i.e.} not the number theoretic) properties of the toral automorphisms
together with  the exponential instability of the dynamics and therefore work
in all dimensions.
\end{abstract}
\thispagestyle{empty}

\section{Introduction}\label{sec:intro}

The Schnirelman theorem states that if a quantum system has  an ergodic
classical limit, then ``most'' of its eigenfunctions equidistribute (on the
energy surface) in phase space in the semi-classical limit. This result has
been proven in many different contexts: for the Laplace-Beltrami operator on
compact Riemannian manifolds with an ergodic geodesic flow in \cite{sc}
\cite{z1} \cite{cdv}, for ergodic billiards in \cite{gl} \cite{zz}, for
non-relativistic quantum mechanics in the classical limit in \cite{hmr}, for
quantum maps in \cite{bodb2} \cite{z2}. A precise statement in the latter
context will be given below (Theorem \ref{thm:schnirel}).

The theorem raises  obvious questions: do there exist exceptional sequences of
eigenfunctions allowing no semi-classical limiting measure or a  limit
different from Liouville measure? It is well known that the limit must in that
case be an invariant probability measure of the dynamics. Clearly, one would
like to better characterize the class of invariant measures that are obtained
as limit measures from sequences of eigenfunctions. Particularly simple
candidates are delta measures concentrated on the periodic orbits of the
dynamics and (finite) convex combinations thereof.  Numerical and theoretical
investigations for ergodic billiards and for quantum maps \cite{bog} \cite{he}
 suggest the possibility that there exist sequences of eigenfunctions
concentrating {\em to some degree} on (unstable) periodic orbits. This
imperfectly defined enhancement phenomenon is loosely referred to as
``scarring''. It is  not clear from the available evidence whether some
sequences of eigenfunctions may concentrate {\em sufficiently strongly} on one
or more periodic orbits to lead to a limiting Dirac measure on those orbits:
no such example is known to date and many researchers in the field seem to
think this should not be possible.  In \cite{cdv}, for example, it is
conjectured that such sequences should not exist on constant negative
curvature surfaces. Partial results in this direction have been obtained using
number theoretic methods for  certain arithmetic hyperbolic surfaces \cite{rs}
\cite{ls} \cite{s3}  (see \cite{s1}, \cite{s2} for a review).

In this paper we analyze the above problem for a simpler class of models that
has attracted much attention, namely the quantized ergodic automorphisms of
the $2d$-torus.   We prove here that for these models such sequences do not
exist (Theorem \ref{thm:noscars}). We also obtain a stronger result that
controls the pure point component of the limiting measures and thereby limits
the class of limit measures (Theorem \ref{thm:noscars2}). Our proofs are based
on an intuitively clear argument that combines the use of the exponential
instability common to all ergodic toral automorphisms (whether they are Anosov
or not) with the algebraic properties of those maps and some basic
semi-classical analysis. They have a distinct dynamical flavour and work in
all dimensions. To put our result in perspective, we will review the
previously known results for the case $d=1$ below. In that case the ergodic
automorphisms are all Anosov and are often referred to as ``cat maps''.

We now describe our results in detail. Unfamiliar concepts and notations are
explained in Section \ref{sec:prepare}. Let $\tor=\R^{2d}/\Z^{2d}$ be the
$d$-dimensional torus, viewed as a symplectic manifold with the canonical
two-form inherited from $\R^{2d}$. Let $A$ be a symplectic and ergodic toral
automorphism, {\em i.e.} $A$ is a symplectic $2d\times 2d$-matrix with integer
entries none of whose eigenvalues are roots of unity. It is known that in that
case at least one of its eigenvalues lies outside the unit cercle so that each
rational point on the torus is an unstable periodic point for $A$. Given such
a periodic orbit $\tau=\{ x_1, x_2,\dots, x_{T_\tau}\}$, we define the delta
measure
$$
\mu_\tau = \frac{1}{T_\tau} \sum_{i=1}^{T_\tau}\delta_{x_i},
$$
which is of course an $A$-invariant measure on $\tor$. Given a finite family
${\cal C}=\cup_{i=1}^K\tau_i$ of periodic orbits, we will also consider the
measures
\begin{equation}\label{eq:dirac}
\mu_{{\cal C},\alpha} = \sum_{j=1}^{K} \alpha_j \mu_{\tau_j}, \quad
\sum_{j=1}^{K}\alpha_j=1,\quad 0\leq \alpha_j\leq 1,
\end{equation}
which are finite convex combinations of the previous ones. These are all
$A$-invariant pure point measures with discrete ({\em i.e.}) finite support.
All invariant Radon measures are obtained by taking the weak closure of those
\cite{ma}. Let $M(A)$ be the unitary quantization of $A$, acting on the
$N^d$-dimensional Hilbert space $\hh$, as defined in Section \ref{sec:prepare}
(We suppress the $N$ and $\kappa$ dependence of $M(A)$ in the notations). Our
main results are the following.
%%%%%%%%%%%%%%%%%%%%%%%%%%%%%%%%%%%%%%%%%%%%%%%%%%%%%%
%%%%%%%%%%%%%%%%%%%%%%%%%%%%%%%%%%%%%%%%%%%%%%%%%%%%%%
\begin{theorem}\label{thm:noscars} Let $A$ be an ergodic symplectic toral automorphism and
let $\psi_N\in\hh, N\in\N$ be  a family of normalized eigenfunctions of
$M(A)$. If the Wigner functions $W_N$ of the $\psi_N$ converge weakly to some
measure $\mu$ on $\T^d$, then $\mu\not=\mu_{{\cal C},\alpha}$, for any choice
of ${\cal C}$ and $\alpha$. In other words, the $W_N$ can not converge to a
pure point measure with discrete support. The same is true for the Husimi
functions $h_N$ of the $\psi_N$.
\end{theorem}
%%%%%%%%%%%%%%%%%%%%%%%%%%%%%%%%%%%%%%%%%%%%%%%%%%%%%%
%%%%%%%%%%%%%%%%%%%%%%%%%%%%%%%%%%%%%%%%%%%%%%%%%%%%%%
This result can be paraphrased by saying that the eigenfunctions can not
concentrate semi-classically  on a finite number of periodic orbits. It is a
particular case of the following more general result.
\begin{theorem}\label{thm:noscars2} Let $A$ be an ergodic symplectic toral automorphism
 and let $\psi_N\in\hh, N\in\N$ be  a family of normalized eigenfunctions of $M(A)$.
 Suppose $\nu$ is a continuous, $A$-invariant probability measure on $\tor$ such that
 for some $0\leq\beta\leq 1$ and for all $f\in C^\infty(\tor)$
\begin{equation}\label{eq:sc}
\lim_{N\to\infty} \langle\psi_{N}\vert Op^W\ f\ \psi_{N}\rangle_{\hh} =
\beta\mu_{{\cal C},\alpha}(f)+(1-\beta)\nu(f)\equiv \mu(f)
\end{equation}
for some ${\cal C}$ and $\alpha$. Then $0\leq\beta\leq (1-\beta)^{1/2}$ or,
equivalently, $\beta \leq (\sqrt 5-1)/2\sim 0,62$.
\end{theorem}

Here $Op^W f$ stands for the Weyl quantization of $f$. Since (\ref{eq:sc}) is
equivalent to the same statement with Weyl quantization replaced by anti-Wick
quantization (see Section \ref{sec:prepare})  it is easy to see that
(\ref{eq:sc}) is equivalent to saying that the absolutely continuous measures
$\mu_N=h_N(x)dx$ converge weakly to $\mu$.  Here $h_N$ is the Husimi function
of $\psi_N$ (See Section \ref{sec:prepare} for a precise definition).  The
result can therefore  loosely be rephrased as follows.
\begin{quote}
If the pure point component of the limiting measure $\mu$ is concentrated on a
{\em finite} number of periodic orbits, then its mass is less than two thirds
of the total mass.
\end{quote}

In \cite{rs}, a sequence $\psi_N$ is defined to ``scar strongly on $\cal C$''
if (\ref{eq:sc}) holds with $\nu$ given by Lebesgue measure. Theorem
\ref{thm:noscars2} does not rule out the possibility of strong scarring, but
limits the size of the scar. In fact, strong scarring does occur in the
systems considered. Indeed, it is proven in \cite{dbfn} that, for $d=1$, there
exists a sequence $N_k\to\infty$, so that for each choice of $\cal C$ and
$\alpha$ as above, there exists a sequence of eigenfunctions $\psi_{N_k}$ so
that
\begin{equation}\label{eq:strongscar}
\lim_{k\to\infty} \langle\psi_{N_k}\vert Op^W\ f\ \psi_{N_k}\rangle_{\hh} =
\frac{1}{2}\mu_{{\cal C},\alpha}(f)+\frac{1}{2}\int_{\T^2} f(x) dx.
\end{equation}
This also shows that, whereas the upper bound on $\beta$ in the theorem is
probably not optimal, one can not do better than $\beta\leq 1/2$.

Theorems \ref{thm:noscars}--\ref{thm:noscars2} can be seen as  partial results
on the characterization of the  measures obtained in the semi-classical limit
from the eigenfunctions of quantized ergodic symplectic automorphisms of
$\tor$. Such measures are sometimes called ``quantum limits''.  As such these
results are to be compared with previous ones for the two-torus available in
the literature. Let us first recall the precise statement of the Schnirelman
theorem for ergodic symplectic toral automorphisms \cite{bodb2}.
\begin{theorem}\label{thm:schnirel} Let $A$, $M(A)$ be as above. Let, for each $N$, $\{\psi_1, \psi_2, \dots
\psi_{N^d}\}$ be a basis of eigenfunctions of $M(A)$. Then, for each $N\in\N$,
there exists a subset $E(N)\subset \{1,\dots, N^d\}$ such that:

(i) $\displaystyle\lim_{N\to\infty} \frac{\sharp E(N)}{N^d}=1;$

(ii) For any $f\in C^\infty(\tor)$, for any sequence $(j_N\in E(N))_{N\in\N}$,
one has
\begin{equation}\label{eq:seqequidistribute}
\lim_{N\to\infty} \langle\psi_{j_N}^{(N)}\vert Op^W\ f\
\psi_{j_N}^{(N)}\rangle_{\hh} = \int_{\tor} f(x) dx.
\end{equation}
\end{theorem}
The strongest possible statement improving on this that one may {\em a priori}
have hoped to prove is this:

\medskip
{\em Let, for each $N\in\N, \psi_N\in \hh$ be a normalized eigenfunction of
$M(A)$. Then, for each $f\in C^\infty(\tor)$, one has}
\begin{equation}\label{eq:equidistributeall}
\lim_{N\to\infty} \langle \psi_N\vert\opwk{f} \psi_N\rangle_{\hh}=\int_{\tor} f(x)
dx.
\end{equation}

\medskip
This is sometimes referred to (in what we feel is a somewhat unfortunate
terminology) as ``quantum unique ergodicity'' and -- as already pointed out --
has not been proven in any chaotic system. Of course, in view of
(\ref{eq:strongscar}) it is obviously not true in the present context of
ergodic toral automorphisms. One nevertheless expects the sequences that
satisfy (\ref{eq:strongscar}) to be rather exceptional: there do indeed exist
 two results in the  direction of (\ref{eq:equidistributeall}), valid for a particular but
 rather large class of
hyperbolic toral automorphisms in $d=1$. The first one is this.
\begin{theorem} \label{thm:hecke} \cite{kuru1} If $A\in$ SL$(2,\Z)$ is hyperbolic and $A\equiv {\mathbb
I}_2$ {\rm mod} $4$, then there exists for each $N$ a basis $\{\psi_1, \psi_2,
\dots \psi_N\}$ of eigenfunctions of $M(A)$ so that
(\ref{eq:seqequidistribute}) holds with $E(N)=\{1,\dots, N\}$.
\end{theorem}
This obviously constitutes a strengthening of the Schnirelman theorem for the
particular class of $A$ considered. The basis for which the result holds is
explicitly described in \cite{kuru1}. Note  the difference between Theorem
\ref{thm:hecke} and (\ref{eq:equidistributeall}). Indeed, the eigenvalues of
$M(A)$ may be degenerate so that  it is possible that exceptional sequences of
eigenfunctions not belonging to the above basis have a different
semi-classical limit. This is all the more true since there exists a sequence
$N_k$ for which the eigenspaces have a $N_k/\ln N_k$ fold degeneracy
\cite{bondb}. It is precisely this sequence of $N$ that is used to construct
the sequence of eigenfunctions in
 (\ref{eq:strongscar}). Another result in
the direction of (\ref{eq:equidistributeall}) is the following.

\begin{theorem} \label{thm:mostN} \cite{kuru2} If $A\in$ SL$(2,\Z)$ is hyperbolic and $a_{11}a_{12}\equiv
0\equiv a_{21}a_{22}$ {\rm mod} $2$, then there exists a density one sequence
of integers $(N_\ell)_{\ell\in\N}$ along which (\ref{eq:equidistributeall})
holds.
\end{theorem}
Theorem \ref{thm:mostN} states that ``quantum unique ergodicity'' holds along
a subsequence of values of $N$.  It is furthermore shown in \cite{kuru2} that,
along this sequence, the degeneracies of the eigenspaces grow sufficiently
slowly  so that it is disjoint from the sequence $N_k$ mentioned above, as it
should be in order not to contradict (\ref{eq:strongscar}). It is therefore
seen  in both theorems that the obstacle to the validity of
(\ref{eq:equidistributeall}) for all $N$ is the existence of growing
degeneracies of the eigenspaces of $M(A)$ for large $N$, as expected.

Our result in Theorem \ref{thm:noscars2} is of a somewhat different nature
than Theorems \ref{thm:hecke} and \ref{thm:mostN}. For any  $A$, $d$ or $N$ it
restricts the candidate limit measures to those that have a ``not too large''
pure point component. In particular, it completely rules out the most
``obvious'' candidates, namely pure point measures with discrete support. Our
result therefore shows in particular that even the very high degeneracies of
the sequence $N_k$ can not be exploited to construct eigenfunctions that
concentrate {\it completely} on unstable periodic orbits.

Quantum mechanics on the torus is usually studied only in the case $d=1$.
Different people have different reasons for imposing this restriction. First,
when doing numerics, higher dimensions quickly poses practical problems of
storage size and computation speed since the dimension of the Hilbert spaces
grows as $\hbar^{-d}$. Next, on the theoretical side, the Schnirelman theorem
is obviously true in all dimensions $d$ since it is proven with
dimension-independent arguments exploiting only the ergodicity of the
dynamics, so there is nothing to be gained from introducing the notational
complications associated with the higher $d$  problems. To prove sharper
results, however, one needs to exploit finer properties of the classical maps.
The proofs of Theorems  \ref{thm:hecke} and \ref{thm:mostN} in \cite{kuru1}
\cite{kuru2} do this by exploiting detailed number theoretic properties of a
particular class of hyperbolic automorphisms of the two-torus and do therefore
{\em not}  carry over in any obvious way to higher dimensions or to general
ergodic symplectic toral automorphisms. In order to stress that the sharpening
of the Schnirelman theorem proven in this paper (Theorems
\ref{thm:noscars}-\ref{thm:noscars2}) exploits only the exponential
instability shared by {\em all} ergodic automorphisms of the $2d$-torus (even
if they are not hyperbolic), as well as their algebraic structure, we have
chosen to consider in the following the general case throughout.

The paper is organized as follows. In section \ref{sec:prepare} we describe
quantum mechanics on the $2d$-torus, and the quantization of symplectic toral
automorphisms, following \cite{bodb2}. We will be as brief as possible,
referring to \cite{bodb2} \cite{bondb} \cite{db2} and references therein for
further information and motivation. In section \ref{sec:ergautom} we recall
some basic facts on ergodic symplectic automorphisms of the $2d$-torus.
Section \ref{sec:proof} is devoted to the  proof of Theorem
\ref{thm:noscars2}. In section \ref{sec:proofquick}, finally,  we  give an
alternative proof of Theorem \ref{thm:noscars} valid only for the case $d=1$ and for a subclass of ergodic automorphisms when $d>1$.
It is based on  a result on the propagation of initially localized states
(Theorem \ref{thm:dynamics}) that generalizes a result of \cite{bondb}. We feel
this result is of interest on its own and in addition it clearly brings out
the central dynamical idea underlying all the results of this paper.

%%%%%%%%%%%%%%%%%%%%%%%%%%%%%%%%%%%%%%%%%%%%%%%%%%%%%%
%%%%%%%%%%%%%%%%%%%%%%%%%%%%%%%%%%%%%%%%%%%%%%%%%%%%%%
%%%%%%%%%%%%%%%%%%%%%%%%%%%%%%%%%%%%%%%%%%%%%%%%%%%%%%

\section{Quantum mechanics on the $2d$-torus}\label{sec:prepare}
In this section we will recall standard facts about quantum mechanics on the
2d--torus $\tor=\R^{2d}/\Z^{2d}$  as well as the quantization of the
symplectic toral automorphisms which was first performed in \cite{hb}. Further
background and references, as well as  proofs, which are omitted here, can be
found in \cite{bondb, bodb2}.

We shall write indifferently $x=(q,p)\in\R^{2d}$ or
$x=(q,p)\in\tor$, where in the latter case $q,p\in [0,1[^d$. Let
$a\cdot b=\sum_i a^ib^i$, for $a,b\in R^d$ and let $\langle (q,p),
(q',p')\rangle = q\cdot p'-q'\cdot p$ be the symplectic form on
$\R^{2d}$. Let $U(a)=\exp-\frac{i}{2\hbar}\langle a,X\rangle$, for
$a\in\R^{2d}$ and $X=(Q,P)$, be the usual representation of the
Heisenberg group on $L^2(\R^d)$, where
$$
(Q_j\psi)(y)=y_j\psi(y),\qquad (P_j\psi)(y)=\frac{\hbar}{i}
\frac{\partial\psi}{\partial y_j}(y).
$$
Let $n=(n_1,n_2)\in\Z^{d}\times\Z^d$ and
$\kappa=(\kappa_2,\kappa_1)\in\R^{2d}/2\pi Z^{2d}$ and let us
define
$$ \hh = \left\{ \psi\in{\cal S}'(\R^d)\ \mid \ U(n)\psi =
\exp \frac{i}{2\hbar}n_1\cdot n_2 \exp i\langle \kappa,n\rangle \psi\right\}\
.$$

\medskip
\begin{lemma}
We have  $\hh\not= \{\emptyset\}$ iff $\exists N\in\N^*$ such that
$(2\pi\hbar) N=1$, in which case ${\rm dim} \hh = N^d$. Moreover,
in that case, $U(n/N)\hh=\hh$ for all $n\in\Z^{2d}$ and there is a
unique Hilbert space structure such that $U(n/N)$ is unitary for
each $n\in\Z^{2d}$.
\end{lemma}
We shall not introduce a different notation for the restriction of
$U(n/N)$ to $\hh$ and in particular not indicate its
$\kappa-$dependence. If $\phi,\psi\in\hh$, we shall write $\langle
\phi\vert \psi\rangle_{\hh}$ or simply $\langle \phi\vert
\psi\rangle$ for their inner product.

We then define  Weyl quantization of a $C^\infty$ function
$f(x)=\sum_{n\in\Z^{2d}}f_n e^{-i2\pi\langle n,x\rangle}$ as
$$
\opwk{f} = \sum_{n\in\Z^{2d}} f_n U(\frac{n}{N}) \;.
$$
Recall that the map
$$
S(\kappa)= (\sum_m \exp -i\kappa_2\cdot m \ U(0,m))(\sum_n\exp i\kappa_1\cdot
n\ U(n,0))
$$
defines a surjection of the space of  Schwartz functions ${\cal S}(\R^d)$ onto
$\hh\subset {\cal S}'(\R^d)$.

Let $\eta_x\in L^2(\R^d)$ denote the usual gaussian Weyl--Heisenberg
coherent state centered on $x\in\R^{2d}$:
$$
\eta_0(y)
=\frac{1}{(\pi\hbar)^\frac{d}{4}}\e{-\frac{1}{2\hbar
}y^{2}},\quad
\eta_{x}(y)=(U(x)\eta_{0})\, (y).
$$
We then define coherent states on the
torus as
\begin{equation}\label{dirintcs}
\eta_{x,\kappa} \equiv
 S(\kappa)\eta_x\in\hh.
\end{equation}
We will find it convenient to use the physicists' ``bra-ket'' notation and to write:
$$
\vert x\rangle \equiv \eta_x\in L^2(\R^d) \quad{\rm and}\quad \vert x,\kappa\rangle \equiv \eta_{x,\kappa}\in\hh.
$$
In particular, we use the notation $\vert x,\kappa\rangle\langle x,\kappa\vert$ to designate the rank one operator associated to $\vert x,
\kappa\rangle$. Coherent states on the torus satisfy the following resolution of the identity
\begin{equation}
\label{res_id} \hbox{Id}_{\hh} =\int_{\tor} \
|x,\kappa\rangle\langle x, \kappa| \frac{dx}{(2\pi\hbar)^d}
\end{equation}
and permit us to define the anti-Wick quantization $\opawk{f}$ of $f\in
L^\infty(\tor)$ as the operator on $\hh$ defined by
\begin{equation}
\opawk{f} =\int_{\tor} f(x)\ |x,\kappa\rangle\langle x,\kappa|
\frac{dx}{(2\pi\hbar)^d}.
\end{equation}
For each $\psi\in\hh$ we define its Wigner function as the distribution
$W_\psi(x)$ such that
$$
\langle \psi\vert \opwk{f}\psi\rangle = \int_\tor f(x) W_\psi(x) dx ~~~\forall
f\in C^\infty(\tor)
$$
and its Husimi function
\begin{equation}\label{eq:husimi}
h_\psi(x)=N|\langle \psi|x,\kappa\rangle |^2
\end{equation}
which satisfies
$$
\langle \psi\vert \opawk{f}\psi \rangle= \int_\tor f(x) h_\psi(x) dx ~~~\forall
f\in L^\infty(\tor)\ .
$$
Anti-Wick and Weyl quantization satisfy for each $f\in C^\infty(\tor)$ the
following estimate:
\begin{equation} \label{eq:waw}
\norm{\opawk{f}-\opwk{f}}\leq \frac{C_f}{N},
\end{equation}
for some positive constant $C_f\geq 0$. Moreover
\begin{equation}\label{eq:shiftcohstate}
U(\frac{n}{N})|x,\kappa\rangle = e^{i\pi\langle n,x\rangle }
|x+\frac{n}{N},\kappa\rangle\; .
\end{equation}

Finally let us come to the quantization of an ergodic symplectic toral
automorphism defined by a matrix $A\in Sp(d,\Z)$. The metaplectic
representation of $Sp(d,\R)$ defines for each $A\in Sp(d, \R)$ a unitary
propagator $M(A)$ in $L^2(\R^d)$ (see \cite{f}); up to a phase it is the
unique operator which satisfies
\begin{equation}
\label{egorov} M(A)^{-1} U(a) M(A) = U(A^{-1}a)\quad \forall a\in\R^{2d}\ .
\end{equation}
The quantization of $A\in Sp(d,\Z)$ on the torus is then straightforward:

\medskip
\begin{lemma}
For each ergodic $A\in Sp(d,\Z)$ and each $N\in \N^*$ there exists at least
one $\kappa\in\R^{2d}/\Z^{2d}$ such that
$$
M(A) \hh = \hh  \;.
$$
\end{lemma}
\dim By applying (\ref{egorov}) one can see that there exists $\kappa_A\in
\R^{2d}/(2\pi\Z)^{2d}$ such that, for each $n\in \Z^{2d}$ and $\psi\in\hh$,
$$
U(n)M(A)\psi = e^{i\pi N A^{-1}n_1\cdot A^{-1}n_2} e^{i\langle
A\kappa,n\rangle} M(A)\psi=e^{i\pi N n_1\cdot n_2} e^{i\langle
\kappa_A,n\rangle} M(A)\psi\;.
$$
As a result $M(A)\hh={\cal H}_\hbar(\kappa_A)$. Indeed let us write
$A^{-1}=(\alpha_{ij})_{ij=1,2}$ with $\alpha_{ij}$ $d\times d$ matrices. Since
$A^{-1}$ is symplectic we have $\alpha_{11}^*\alpha_{22}-
\alpha_{21}^*\alpha_{12}=1$,
$\alpha_{11}^*\alpha_{21}=\alpha_{21}^*\alpha_{11}$ and
$\alpha_{12}^*\alpha_{22}=\alpha_{22}^*\alpha_{12}$. Consequently a simple
computation shows that
$$
A^{-1}n_1 \cdot A^{-1}n_2-n_1\cdot n_2 = \langle \omega_A,n\rangle ~~~~~{\rm
mod}\ 2
$$
where
$$
\omega_A = \left(\begin{array}{c}{\rm diag}(\alpha_{12}^*\alpha_{22})\cr {\rm
diag}(\alpha_{11}^*\alpha_{21})\end{array}\right)
$$
and
\begin{equation}\label{kappa}
\kappa_A = A \kappa + \pi N \omega_A  ~~{\rm mod}\ 2\pi~.
\end{equation}
Since $A$ is ergodic, $1$ is not an eigenvalue of $A$ and equation
(\ref{kappa}) admits at least one fixed point $\kappa_A=\kappa$. \endproof

\medskip
In the following, we shall always assume that $\kappa$ has been chosen as in the Lemma, but we shall not explicitly indicate the $N$ or $A$ dependence of $\kappa$. Similarly, we shall use the  symbol $M(A)$ to indicate the restriction of $M(A)$ to
$\hk$ for a suitable $\kappa$ as above, without indicating its $N$ or $\kappa$ dependence.

From this construction it follows easily that,
for each $f\in C^\infty(\tor)$,
\begin{equation}\label{eq:egorov}
M(A)^{-1} Op_\kappa^W f M(A) = Op_\kappa^W (f\circ A),\ \forall f\in
C^\infty(\tor).
\end{equation}
In other words  ``quantization and  evolution commute''.
\bigskip

%%%%%%%%%%%%%%%%%%%%%%%%%%%%%%%%%%%%%%%%%%%%%%%%%%%%%%
%%%%%%%%%%%%%%%%%%%%%%%%%%%%%%%%%%%%%%%%%%%%%%%%%%%%%%
%%%%%%%%%%%%%%%%%%%%%%%%%%%%%%%%%%%%%%%%%%%%%%%%%%%%%%
\section{Ergodic automorphisms of the torus}\label{sec:ergautom}
We collect here some rather basic facts about ergodic automorphisms of the
torus. Let $A\in$ SL$(2d,\Z)$, then $A$ defines an ergodic toral automorphism
if and only if none of the eigenvalues of $A$ are roots of unity \cite{m}.
Ergodic toral automorphisms are automatically mixing as well \cite{m}. In
addition, their eigenvalues can not all lie on the unit circle: at least one
of them has to have a modulus strictly bigger than $1$. This is an immediate
consequence of the Kronecker theorem (\cite{n}, Theorem 2.1), applied to the
characteristic polynomial of $A$.  As a result, in the decomposition of
$\R^{2d}$ into $A$-invariant subspaces \cite{kh} \cite{li} given by
$$
\R^{2d}= E_- \oplus E_0 \oplus E_+,
$$
where $E_+$ (respectively $E_0, E_-$) is the root space of $A$ corresponding
to eigenvalues of modulus strictly bigger than (respectively equal to,
strictly smaller than) $1$, we are sure that $E_-$, $E_+$ are non-trivial. A
matrix $A$ is said to be hyperbolic iff $E_0=\{0\}$. The corresponding
dynamical system on $\tor$ is then Anosov. If $E_0\not=\{0\}$, $A$ is called
quasi-hyperbolic in \cite{li}.  Clearly, when $d=1$, all ergodic toral
automorphisms are Anosov, but this is no longer true in higher dimension.

%Let $E_\lambda$ be the eigenspace corresponding to the eigenvalue
%$\lambda$, $n_\lambda={\rm dim}\ E_\lambda$ and let
%$\R^{2d}=\oplus_\lambda E_\lambda$. A simple application of the
%Jordan decomposition of $A$ gives (\cite{ka}) for $x=\sum_\lambda
%x_\lambda$
%\begin{equation}
%\label{jordan}
%C_-t^{-(n_{\lambda_+}-1)}|\lambda_+|^t\norm{x_{\lambda_+}}\leq\norm{A^tx}\leq
%C_+ t^{n_{\lambda_+}-1} |\lambda_+|^t \norm{x_{\lambda_+}}
%\end{equation}
%where $\lambda_+$ is the biggest eigenvalue such that
%$x_{\lambda_+}\not=0$.

We will need the following result:
\begin{lemma}\label{lem:expgrowth} Let $n\in\Z^{2d}$. Then $n\not\in E_0\oplus E_-$.
Moreover, there exist $\gamma>0, C_\pm>0$ and $0\leq k\leq 2d-2$ so that, for
all $t\in\N$ large enough
$$
C_- t^{k} e^{\gamma t} \leq \parallel A^tn\parallel \leq C_+ t^{k} e^{\gamma
t}.
$$
\end{lemma}
\noindent{\bf Proof:} The first statement, namely that $E_0\oplus E_-\cap
\Z^{2d}=\{0\}$,
 can be found in \cite{ka}\cite{li}.
The estimate is then a  simple application of the Jordan normal
form.\endproof It should be noted that $\gamma>0, C_\pm>0$ and
$0\leq k\leq 2d-2$ depend on $n$ in the above estimate.  The lower
bound above is an expression of the exponential instability common
to all ergodic toral automorphisms and is the only information
about them we shall need to prove Theorems \ref{thm:noscars} and
\ref{thm:noscars2}.

To prove Theorem \ref{thm:dynamics} however, we will need the following
result from \cite{ka}, which is the generalization to higher dimensions of the obvious diophantine inequality satisfied in the case $d=1$ by the slopes of the stable and unstable directions of $A$.

\begin{lemma}
\label{lemma_ka} Let $\R^{2d}=V_1\oplus V_2$, with $V_i$ invariant
spaces for $A$ such that
\begin{itemize}
\item[{\it i)}] $A |_{V_1}$ and $A|_{V_2}$ don't have common
eigenvalues;
\item[{\it ii)}] $V_1\cap \Z^{2d}=\{0\}$.
\end{itemize}
Then there exists $C>0$ such that for each $n\in\Z^{2d}_*$ we have
($m={\rm dim} V_1$)
$$ d_{\R^{2d}}(n,V_1) \geq
\frac{C}{\norm{n}^{m}} \;.
$$
\end{lemma}
%To prove Theorem \ref{thm:dynamics}, we need a more detailed lower bound. When
%$d=1$, the stable and unstable directions $v_\pm$ of $A$ have  slopes $s_\pm$
% that are
%quadratic irrationals, as is easily checked. The following diophantine
%inequality is then a basic property  of all quadratic irrationals \cite{kh}:
%\begin{equation}\label{eq:dioph}
%\exists d_\pm>0, \forall a\in\Z, b\in\Z_*, \qquad |bs_\pm - a|\geq
%\frac{d_\pm}{|b|}.
%\end{equation}
%Writing $e^\gamma, \gamma>0$ for the largest eigenvalue of $A$,  one then
%easily finds:
%\begin{equation}\label{eq:diophantine}
%\exists C_-, C_+>0, \forall n\in\Z^2_*, t\in \N, \qquad C_- e^{\gamma
%t}\frac{1}{\parallel n\parallel}\leq \parallel A^tn\parallel\leq C_+e^{\gamma
%t}\parallel n\parallel.
%\end{equation}

%%%%%%%%%%%%%%%%%%%%%%%%%%%%%%%%%%%%%%%%%%%%%%%%%%%%%%%%%%
%%%%%%%%%%%%%%%%%%%%%%%%%%%%%%%%%%%%%%%%%%%%%%%%%%%%%%%%%%5
\section{Proof of Theorem \ref{thm:noscars2}}\label{sec:proof}
We will need the following simple technical lemma.
\begin{lemma}\label{lem:smoothchar}
Let $B$ be a Borel subset of $\tor$. Then, under the hypotheses of Theorem
\ref{thm:noscars2}, one has
\begin{equation}\label{eq:smoothcar}
\mu(\hbox{\rm int } B)\leq \liminf_{N\to+\infty}\langle \psi_N\vert
\opawk{\chi_B}\psi_N\rangle\leq\limsup_{N\to+\infty}\langle \psi_N\vert
\opawk{\chi_B}\psi_N\rangle \leq \mu(\bar B).
\end{equation}
where $\chi_B$ is the characteristic function of $B$.
\end{lemma}
\noindent{\bf Proof:} This is a standard result in measure theory, we include
the proof for completeness. Let us introduce, for all $\epsilon>0$,
$$
B_-^\epsilon = \{x\in B | d_{\tor}(x, \partial B)>\epsilon\}\subset B \subset
B_+^\epsilon = \{x\in \tor | d_{\tor}(x, B)<\epsilon\},
$$
where $d_{\tor}$ designates the Euclidean distance on the torus. Then we have
$ \cup_{\epsilon>0} B_-^\epsilon = \hbox{\rm int }B\ \hbox{\rm and }\
\cap_{\epsilon>0} B_+^\epsilon = \bar B, $ so that
\begin{equation}\label{eq:limeps}
\lim_{\epsilon\to0} \mu(B_-^{\epsilon}) = \mu(\hbox{\rm int } B)\qquad
\lim_{\epsilon\to0} \mu(B_+^{\epsilon}) = \mu(\bar B).
\end{equation}
Now let $\eta\in C^\infty_0(\R^{2d})$ be a spherically symmetric positive
function, with support in the ball of radius $1$, equal to $1$ on the ball of
radius $1/2$ and such that $\int \eta(x) dx = 1$. We set $ \eta_\epsilon(x)=
\frac{1}{\epsilon^{2d}}\eta(\frac{x}{\epsilon}), $ and define $
\chi_{\pm}^\epsilon(y)  = \int_{B_{\pm}^\epsilon} \eta_\epsilon(y-x) dx. $
Clearly
\begin{eqnarray*}
\chi_+^\epsilon(y)  = 1\ \hbox{\rm if }\ y\in B&\ &
\chi_+^\epsilon(y)  = 0\ \hbox{\rm if }\ y\in \tor\setminus B_+^{2\epsilon}\\
\chi_-^\epsilon(y)  = 1\ \hbox{\rm if }\ y\in B_-^{2\epsilon}&\ &
\chi_-^\epsilon(y)  =0\  \hbox{\rm if }\ y \in \tor\setminus B.
\end{eqnarray*}
This implies in particular that $ \chi_-^\epsilon\leq \chi_B\leq
\chi_+^\epsilon, $ so that the positivity of anti-Wick quantization implies
that $ \opawk{\chi_-^\epsilon}\leq \opawk{\chi_B}\leq \opawk{\chi_+^\epsilon}.
$ Using (\ref{eq:sc}) and (\ref{eq:waw}) we then find
\begin{eqnarray*}
\mu(B_-^{2\epsilon})\leq\mu(\chi_-^\epsilon)&\leq&
\liminf_{N\to+\infty}\langle\psi_N\vert\opawk{\chi_B}\psi_N\rangle\\
&\leq&
\limsup_{N\to+\infty}\langle\psi_N\vert\opawk{\chi_B}\psi_N\rangle
\leq \mu(\chi_+^\epsilon)\leq
\mu(B_+^{2\epsilon})
\end{eqnarray*}
so that the result follows by taking $\epsilon\to0$ and using
(\ref{eq:limeps}).\endproof

Define, for any finite set of points $\cal C$ (not necessarily a set of
periodic points of the dynamics) and for each $a>0$
\begin{equation}\label{eq:cutball}
B_a=\{x\in\tor \mid d_{\tor}(x,{\cal C})<a\}.
\end{equation}
We also introduce
\begin{equation}\label{eq:radius}
\delta_{\cal C} = \min \{d_{\tor}(x,y) | x,y\in {\cal C}, x\not= y\},
\end{equation}
provided $\cal C$ contains more than one point. Otherwise we define
$\delta_{\cal C}=1/\sqrt2$.

\noindent{\bf Proof of Theorem \ref{thm:noscars2}:}  Since $\nu$ is a
continuous measure, the Wiener theorem says that
$$
\lim_{K\to+\infty}\frac{1}{(2K+1)^d}\sum_{\parallel n\parallel \leq K} \vert \hat \nu (n)\vert^2 =0.
$$
Here, $\hat \nu$ is the Fourier transform of $\nu$. This implies immediately that there exists a density one
subset $G$ of $\Z^{2d}$ so that
\begin{equation}\label{eq:wiener}
\lim_{n\to \infty, n\in G} \hat \nu(n)=0.
\end{equation}
On the other hand,  $\cal
C$ is a finite collection of rational points on $\tor$ and we call $S$ the
least common multiple of the denominators of those points. Then, for each
$n\in S\Z^{2d}$ and for all $x\in {\cal C}$, clearly $\chi_n(x)\equiv\exp 2\pi
i \langle n, x\rangle=1$ and consequently
$$
\mu_{{\cal C},\alpha}(\chi_n)=1.
$$
Hence, for such $n$,
\begin{equation}\label{eq:limcharacter}
\lim_{N\to\infty}\langle \psi_N\vert U(\frac{n}{N})\psi_N\rangle = \beta
+(1-\beta)\hat \nu (n).
\end{equation}
Since $S\Z^{2d}$ is a positive density subset of $\Z^{2d}$, it follows that
$S\Z^{2d}\cap G$ is positive density as well. As a result, given $\epsilon>0$,
there exists $n\in S\Z^{2d}\cap G$, depending on $\epsilon$, so that
\begin{equation}\label{eq:smallfourier}
|\hat \nu(n)|\leq \epsilon.
\end{equation}
We then have, using respectively (\ref{eq:egorov}), (\ref{res_id}) and (\ref{eq:shiftcohstate})
\begin{eqnarray}\label{eq:2}
\mid\langle \psi_N\vert  U(\frac{n}{N}) \psi_N\rangle\mid &=&
\mid\langle \psi_N\vert M(A)^{-t} U(\frac{n}{N}) M(A)^t\psi_N\rangle\mid\nonumber\\
&=&\mid\langle \psi_N\mid  U(\frac{A^t n}{N}) \psi_N\rangle\mid\nonumber\\
&=&\mid\int_{\tor} \langle \psi_N\mid x,\kappa\rangle\langle x,\kappa\mid U(\frac{A^t n}{N})
 \psi_N\rangle \frac{dx}{(2\pi\hbar)^{d}}\mid\nonumber\\
&\leq&\int_{\tor} \mid\langle \psi_N\mid x,\kappa\rangle\mid\, \mid\langle
x-\frac{A^t n}{N},\kappa\mid
 \psi_N\rangle\mid \frac{dx}{(2\pi\hbar)^{d}}\nonumber\\
&\leq&\int_{B_a} \mid\langle \psi_N\mid x, \kappa\rangle\mid\, \mid\langle x-\frac{A^t
n}{N},\kappa\mid
 \psi_N\rangle\mid \frac{dx}{(2\pi\hbar)^{d}}\nonumber\\
&\ &\qquad\qquad + \bigl(\int_{\tor \setminus B_a} \mid\langle \psi_N\mid
x,\kappa\rangle\mid^2 \frac{dx}{(2\pi\hbar)^{d}}\bigr)^{1/2},
\end{eqnarray}
where $B_a$ is defined in (\ref{eq:cutball}). Note that this inequality holds
for each choice of $t, N, a$. Now choose $M>3$ and such that
\begin{equation}\label{eq:cond1}
\frac{1}{\gamma}(\ln \frac{M}{C_+}-\ln \frac{3}{C_-})>1+\frac{k}{\gamma},
\end{equation}
where $C_{\pm}, k$ and $\gamma$ are defined in Lemma \ref{lem:expgrowth}. We
recall they depend on $n$. We will show below that then, for all $a>0$, the
following is true:
\begin{equation}\label{eq:times}
\forall N>N_a=\frac{C_-e^\gamma}{3}\frac{1}{a}, \exists t_N\in\N \ \hbox{\rm
so that }\ 3a\leq
\parallel \frac{A^{t_N}n}{N}\parallel\leq Ma<\delta_{\cal C}/3.
\end{equation}
Introducing
\begin{equation}\label{eq:annulus}
{\cal A}(a,M)= \{x\in\tor \mid 2a\leq d_{\tor}(x,{\cal C})\leq (M+1)a\},
\end{equation}
it is then clear that
$$
x\in B_a\Rightarrow  x-\frac{A^{t_N} n}{N}\in {\cal A}(a,M).
$$
The important point here is that ${\cal A}(a,M)$ does not depend on $N$.
Inequality (\ref{eq:2}) now yields, upon using a Schwartz inequality in the
first term
\begin{eqnarray}\label{eq:4}
|\langle \psi_N\mid  U(\frac{n}{N})\psi_N\rangle\mid&\leq& \langle
\psi_N\mid \opawk {\chi_{B_a}}\psi_N\rangle^{1/2} \langle
\psi_N\mid \opawk {\chi_{{\cal
A}(a,M)}}\psi_N\rangle^{1/2}\nonumber\\
&\ &\qquad\qquad\qquad+ \langle \psi_N\mid \opawk {\chi_{\tor\setminus
B_a}}\psi_N\rangle^{1/2}\nonumber \\
&\leq&
\langle \psi_N\mid \opawk {\chi_{{\cal A}(a,M)}}\psi_N\rangle^{1/2}\nonumber\\
&\ &\qquad\qquad\qquad+ \langle \psi_N\mid \opawk {\chi_{\tor\setminus
B_a}}\psi_N\rangle^{1/2}.
\end{eqnarray}
Note that, given $\epsilon>0$, this inequality holds for $n$ satisfying
(\ref{eq:smallfourier}), for all $a$ small enough (depending on $n$), and for
all $N$. We now take the limsup for $N$ to $+\infty$, and  apply Lemma
\ref{lem:smoothchar} in the right-hand side and
(\ref{eq:limcharacter})-(\ref{eq:smallfourier}) in the left-hand side to
obtain:
\begin{equation}
\beta -(1-\beta)\epsilon\leq (1-\beta)^{1/2} \nu(\overline{{\cal
A}(a,M)})^{1/2}+ (1-\beta)^{1/2}\nu(\overline{\tor\setminus B_a})^{1/2}.
\end{equation}
  Finally, taking, for $\epsilon$ and $M$ fixed, $a$ to $0$ in this inequality, the continuity of the
measure $\nu$ yields
$$
\beta-(1-\beta)\epsilon\leq (1-\beta)^{1/2}.
$$
Since this holds for all $\epsilon$, this is the desired result.

It remains to prove (\ref{eq:times}). From Lemma \ref{lem:expgrowth} we see
that  (\ref{eq:times}) will be proven provided we show there exists, for each
$N\in\N$, $N\geq N_a$ a $t_N\in\N$ so that
$$
3aN\leq C_- t_N^ke^{\gamma t_N},\qquad C_+ t_N^ke^{\gamma t_N}\leq NMa,
$$
or, equivalently
$$
D_-\equiv \frac{1}{\gamma}[\ln N + \ln a +\ln \frac{3}{C_-}] \leq
t_N+\frac{k}{\gamma}\ln t_N\leq \frac{1}{\gamma}[\ln N + \ln a +\ln
\frac{M}{C_+}]\equiv D_+.
$$
Introducing for $t\in\N_*$, $g(t)=t+\frac{k}{\gamma}\ln t$, one sees that for
all $t\in\N_*$ $g(t+1)-g(t)\leq 1+\frac{k}{\gamma}$. Hence, to obtain
(\ref{eq:times}) it is sufficient that
$$
D_+-D_-=\frac{1}{\gamma}(\ln \frac{M}{C_+}-\ln
\frac{3}{C_-})>1+\frac{k}{\gamma},
$$
and that $D_-\geq g(1)$, but this is guaranteed by condition (\ref{eq:cond1})
and the definition of $N_a$ in (\ref{eq:times}).

\endproof

%%%%%%%%%%%%%%%%%%%%%%%%%%%%%%%%%%%%%%%%%%%%%%%%%%%%%%
%%%%%%%%%%%%%%%%%%%%%%%%%%%%%%%%%%%%%%%%%%%%%%%%%%%%%%
%%%%%%%%%%%%%%%%%%%%%%%%%%%%%%%%%%%%%%%%%%%%%%%%%%%%%%
\section{Propagating localized states}\label{sec:proofquick}
In this section we present a generalization of the main result of \cite{bondb}.
When $d>1$, it only holds under some mild
additional hypotheses on $A$ specified below. Under these conditions, it provides an alternative proof of Theorem \ref{thm:noscars}.  We feel this result is of interest on its own, and in addition  it clearly brings out the basic
``dynamical'' intuition underlying the  proofs of the previous section.

We will impose, in addition to ergodicity, two more conditions on $A$. First, we ask that $A\in $Sp$(d,\Z)$ does not  leave  any non-trivial sublattice of $\Z^{2d}\subset\R^{2d}$ invariant. This excludes, for example, in the case $d=2$, matrices of the form
$$
\left(\begin{array}{cc}
A_{1}&0\\
0&A_{2}
\end{array}\right)
$$
where each bloc $A_1, A_2$ is a hyperbolic matrix in SL$(2,\Z)$.
In addition, we impose the following condition. Let $e^{\gamma_{+}}$
be the maximal modulus of the eigenvalues of $A$ and let $E_{\gamma_+}$ be the corresponding root
space ({\it i.e.}
$E_{\gamma_+}=\oplus_{|\lambda|=e^{\gamma_+}}E_\lambda$);
 we demand that the restriction of $A$ to $E_{\gamma_+}$
is diagonalizable. This will obviously be the case if all roots of the characteristic polynomial of $A$ are distinct, for example. We need some more notations.  Let $m_{\gamma_+}={\rm dim}\ E_{\gamma_+}$
and $m_+={\rm dim} (E_+\oplus E_0)$; of course $1\leq
m_{\gamma_+}\leq m_+\leq d$.

We remark that in $d=1$ any ergodic
matrix $A\in SL(2,\Z)$ satisfies these requirements.

\begin{theorem} \label{thm:dynamics} Let $A$ be as above and let $\mu$ be a pure point
probability measure on $\tor$, with finite support $\cal C$. Let
$\psi_N\in\hh, ||\psi_N||=1$ be a sequence of normalized vectors
in $\hh$ such that, for all $f\in C^\infty(\tor)$
\begin{equation}\label{eq:scar2}
\lim_{N\to\infty} \langle \psi_N\mid\opwk{f} \psi_N\rangle=\mu(f).
\end{equation}
Let $a_N$ be a sequence of positive numbers tending to $0$ with the property
that
\begin{equation}\label{eq:concentrate}
\int_{B_{a_N}} h_N(x) dx\to 1,
\end{equation}
where $B_{a_N}$ is defined in (\ref{eq:cutball}) and $h_N\equiv h_{\psi_N}$ in (\ref{eq:husimi}). Then, there
exist $t_{\pm}\geq 0$ so that, for any sufficiently slowly growing
sequence of integers $\theta_N$ ( i.e.
$\theta_N<\frac{(1-\epsilon)(1+m_+)}{m_+\gamma_+(2d+1-m_{\gamma_+})}\ln
\frac{1}{a_N}$ for some $\epsilon>0$), one has, for each $f\in
C^{\infty}(\tor)$,
\begin{equation}\label{eq:expmixing}
\lim_{N\to\infty} \langle \psi_N \mid M(A)^{-t}\opwk{f}M(A)^{t}
\psi_N\rangle = \int_{\tor} f(x) dx
\end{equation}
uniformly for all $t$ in the region
\begin{equation}\label{eq:region}
\frac{1}{\gamma_+}\ln Na_N + t_-+ (2d-m_{\gamma_+})\theta_N\leq
t\leq \frac{1}{\gamma_+}\ln \frac{N}{a_N^{1/m_+}} - t_+ - \theta_N
\end{equation}
\end{theorem}
Note that here the $\psi_N$ are of course not assumed to be  eigenvectors of
the dynamics. Remark furthermore that the hypothesis (\ref{eq:scar2})
immediately implies the existence of a sequence $a_N$. It is finally clear
from the definition of the Husimi functions in Section \ref{sec:prepare} that
$a_N\sqrt N$ is bounded away from $0$.

As an example, suppose ${\cal C}=\{x_1, x_2, \dots x_p\}$ is a set of $p$
points on the torus and take
$$
\psi_N = \frac{1}{\sqrt p}\sum_{i=1}^p |x_i, \kappa\rangle.
$$
In that case $a_N=N^{-\frac{1}{2}+\epsilon}$ for any $\epsilon$. It is also
easy to check that the $\psi_N$ are normalized up to an exponentially small
factor. The case $p=1$ and $d=1$ was treated in \cite{bondb} but with a worse upper
bound on the times in (\ref{eq:region}).  As explained already in
\cite{bondb}, the  upper bound in Theorem \ref{thm:dynamics} is the optimal one for this case; it is  obtained here
using  an argument borrowed from \cite{dbfn}.

\noindent{\bf Alternative proof of Theorem \ref{thm:noscars} for $A$ as above:} The proof goes by
contradiction. Suppose a sequence of {\em eigenfunctions} exists, so that
(\ref{eq:scar2})  holds with $\mu=\mu_{{\cal C},\alpha}$ (see
(\ref{eq:dirac})). Since  the $\psi_N$ are eigenfunctions, one trivially finds,
for all $t\in \N$:
$$
\langle \psi_N\mid M(A)^{-t}\opwk{f}M(A)^{t} \psi_N\rangle
=\langle \psi_N\mid\opwk{f} \psi_N\rangle,
$$
so that (\ref{eq:expmixing}) implies that
$$
\lim_{N\to\infty} \langle \psi_N\mid \opwk{f}
\psi_N\rangle=\int_{\tor} f(x) dx,
$$
which is in obvious contradiction to the hypothesis (\ref{eq:scar2}). In other
words, we have just proven that, if a sequence of eigenfunctions concentrates
semi-classically on a finite family of periodic orbits, then it
equidistributes. We conclude that such a sequence does not exist.
\endproof

\noindent{\bf Proof of Theorem \ref{thm:dynamics}:} Writing
$$
f=\sum_{\parallel n\parallel \leq M_N} f_n\chi_n + \sum_{\parallel n\parallel
\geq M_N} f_n\chi_n,
$$
the fast decrease of the $f_n$ implies that for all $K\in\N$, there exists
$C_f, C_{f,K}>0$ so that
$$
|\langle \psi_N\mid M(A)^{-t}\opwk{f}M(A)^{t} \psi_N\rangle -
\int_{\tor} f(x) dx|\leq\qquad\qquad\qquad\qquad\qquad
$$
\begin{equation}\label{eq:estim}
\qquad\qquad\qquad C_f\sup_{0<\parallel n\parallel \leq M_N} \mid\langle
\psi_N\mid M(A)^{-t} U(\frac{n}{N}) M(A)^{t}\psi_N\rangle + C_{f,K} M_N^{-K}.
\end{equation}
Hence it will be enough to show that there exist  a sequence $M_N\in\N$ (depending
on $\theta_N$) with $M_N\to+\infty$ so that
\begin{equation*}\label{eq:goal}
\sup_{0<\parallel n\parallel \leq M_N} \mid\langle \psi_N\mid
M(A)^{-t} U(\frac{n}{N}) M(A)^{t}\psi_N\rangle\mid\to 0
\end{equation*}
for $t$ in the range given in (\ref{eq:region}).  For that
purpose, first note that for each $n\in\Z^{2d}\setminus\{0\}$,
$t\in\Z, a>0$, we have (as in (\ref{eq:2}))
\begin{eqnarray}\label{eq:1}
\mid\langle \psi_N\mid M(A)^{-t} U(\frac{n}{N}) M(A)^t\psi_N\rangle\mid &=&
\mid\int_{\tor} \langle \psi_N\mid x,\kappa\rangle\langle x,\kappa\mid U(\frac{A^t n}{N})
 \psi_N\rangle \frac{dx}{(2\pi\hbar)^{d}}\mid\nonumber\\
&\leq&\int_{B_a} \mid\langle \psi_N\mid x,\kappa\rangle\mid\, \mid\langle x+\frac{A^t
n}{N},\kappa\mid
 \psi_N\rangle\mid \frac{dx}{(2\pi\hbar)^{d}}\nonumber\\
&\ &\qquad\qquad\qquad\qquad\qquad\qquad+ \sigma_N(a)\end{eqnarray} where
$$
\sigma_N(a) = \bigl(\int_{\tor \setminus B_a} h_{\psi_N}(x)
dx\bigr)^{1/2}.
$$
The hypothesis (\ref{eq:concentrate}) implies that $\sigma_N(a_N)\to0$.

Below, we shall prove that, for each $N$, there exists $M_N\in\N$ so that
\begin{equation}\label{eq:trick}
x\in B_{a_N}, n\in\Z^{2d}, 0<\parallel n\parallel \leq M_N
\Rightarrow x+\frac{A^{t} n}{N}\not\in B_{a_N},
\end{equation}
for all $t$ in the region (\ref{eq:region}). It then follows from (\ref{eq:1})
and the Schwartz inequality that, for each $\parallel n\parallel \leq M_N$ and
for those $t$
$$
\mid\langle \psi_N\mid M(A)^{-{t}} U(\frac{n}{N}) M(A)^{t}\psi_N\rangle\mid \leq
\sigma_N(a_N) +\sigma_N(a_N).
$$
Hence
\begin{equation}\label{eq:estim2}
|\langle \psi_N\mid M(A)^{-t}\opwk{f}M(A)^{t} \psi_N\rangle -
\int_{\tor} f(x) dx|\leq 2C_f \sigma_N(a_N) + C_{f,K} M_N^{-K}.
\end{equation}

We now prove (\ref{eq:trick}). For simplicity of notation, let us
first consider the case where ${\cal C}=\{0\}$ so that $B_{a_N}$
is the ball of radius $a_N$ around $0\in\T^{2d}$. We define
$$
\tilde B_{a_N} = \{ y\in\R^{2d} | d_{\R^{2d}} (y, \Z^{2d})< a_N\}.
$$
Then $B_{a_N}$ is the image of $\tilde B_{a_N}$ under the natural
projection of $\R^{2d}$ to $\tor$ so that (\ref{eq:trick}) is
equivalent to
$$
x\in \tilde B_{a_N}, n\in\Z^{2d}, 0<\parallel n\parallel \leq M_N
\Rightarrow d_{\R^{2d}}(x+\frac{A^{t} n}{N}, \Z^{2d}) \geq a_N.
$$
But this is guaranteed if for all $t$ in (\ref{eq:region})
\begin{equation}\label{eq:trickbis}
n\in\Z^{2d}, 0<\parallel n\parallel \leq M_N \Rightarrow
d_{\R^{2d}}(\frac{A^{t} n}{N}, \Z^{2d}) \geq 2a_N.
\end{equation}
This is what we now prove. First, let
$$
e^{-\gamma_-}=\max \{|\lambda|< 1 \mid \lambda \ \hbox{\rm is an eigenvalue of }\ A\}.
$$
In other words, $e^{-\gamma_-}$ is the modulus of the largest eigenvalue of $A$ strictly inside the unit disc.
Since $A$ does not leave any non-trivial sublattice invariant, it is clear that  the component along
$E_{\gamma_+}$ of any $n\in\Z^{2d}_*$ is different from zero and Lemma
\ref{lemma_ka} can therefore be used with $V_2=E_{\gamma_+}$; since $A$ is
diagonalizable on $E_{\gamma_+}$, we conclude that there exist
$C_{\pm}>0$ so that for all $n\in\Z^{2d}_*$ and for all $t$ one
has
\begin{equation}\label{eq:dioph2}
\frac{C_-}{\norm{n}^{2d-m_{\gamma_+}}} e^{\gamma_+ t}\leq
\parallel A^t n\parallel \leq C_+ e^{\gamma_+ t} \norm{n}.
\end{equation}
%%%%%%%%%%%%%%%%%%%%%
By using (\ref{eq:dioph2}), if
$$
\frac{C_-}{NM_N^{2d-m_{\gamma_+}}}e^{\gamma_+ t}\geq 2 a_N,
$$
it is clear that for each $n\in\Z^{2d}_*$ such that $\norm{n}\leq
M_N$ we have
$$
\parallel \frac{A^tn}{N}\parallel \geq 2a_N.
$$
So, if we choose $M_N=e^{\gamma_+ \theta_N}, t_-=
\frac{1}{\gamma_+}\ln\frac{2}{C_-}$, this latter inequality is
satisfied for all $t$ in (\ref{eq:region}). We note for later
reference that, since $2d-m_{\gamma_+}\geq m_+$,
\begin{equation}\label{eq:estMN}
M_N\leq
\left(\frac{1}{a_N}\right)^{\frac{1+m_+}{m_+(2d+1-m_{\gamma_+})}}\leq
C N^{1/2}
\end{equation}
for some $C>0$, in view of the constraint on $\theta_N$ and the
fact that $a_N\sqrt N$ is bounded away from $0$. Now, for each
$n\in\Z^{2d}, 0<\parallel n\parallel \leq M_N$ there exists
$n_{t,N}\in\Z^{2d}$ so that
$$
d_{\R^{2d}}(\frac{A^{t} n}{N}, \Z^{2d})= \parallel \frac{A^{t}
n}{N} - n_{t,N}\parallel.
$$
Consequently, if $n_{t,N}=0$ then  $d_{\R^{2d}}(\frac{A^{t} n}{N},
\Z^{2d}) \geq 2a_N$ for all $t$ in the region (\ref{eq:region}).
Suppose therefore that $n_{t,N}\not=0$ so that $\parallel
\frac{A^{t} n}{N}\parallel \geq 1/2$. Let $n=n_{+}+n_0+n_-\in
E_+\oplus E_0\oplus E_-$. Then we have
\begin{eqnarray*}
d_{\R^{2d}}(\frac{A^tn}{N}, \Z^{2d})&=&
\norm{\frac{A^tn}{N}-n_{t,N}}
\geq\norm{\frac{A^t(n_++n_0)}{N}-n_{t,N}}-\norm{\frac{A^tn_-}{N}}\cr
&\geq& d_{\R^{2d}}(n_{t,N},E_+\oplus E_0) -
\norm{\frac{A^tn_-}{N}}\geq
\frac{C_o}{\norm{n_{t,N}}^{m_+}}-\norm{\frac{A^tn_-}{N}}\cr
&\geq& \frac{C_o}{(\sqrt{d/2}+\norm{A^tn/N})^{m_+}}-\norm{\frac{A^tn_-}{N}}\geq
\frac{C_1}{\norm{A^tn/N}^{m_+}}-\norm{\frac{A^tn_-}{N}}\cr
&\geq& C_2\left(\frac{N}{M_N}e^{-\gamma_+t}\right)^{m_+}-
C_3\frac{M_N}{N} t^{(d-1)}e^{-\gamma_-t}\cr
&=& C_2\left(\frac{N}{M_N}e^{-\gamma_+t}\right)^{m_+}
\left[1-\frac{C_3}{C_2}e^{(m_+\gamma_+-\gamma_-)t}
\left(\frac{M_N}{N}\right)^{m_++1}t^{(d-1)}\right]\cr
&\geq& 2 a_N \;,
\end{eqnarray*}
where we used in the second line Lemma \ref{lemma_ka} applied to
$V_1=E_+\oplus E_0$, in the third line the fact that
$\norm{A^tn/N}\geq 1/2$, in the fourth line the upper bound in
(\ref{eq:dioph2}) and a standard estimate on $\norm{A^tn_-}$. To
obtain the last line one defines $t_+$ via
$e^{-\gamma_+t_+m_+}=C_2/4$, one uses $M_N=\exp \gamma_+\theta_N$
and (\ref{eq:region}) to obtain
$$
e^{(m_+\gamma_+-\gamma_-)t}
\left(\frac{M_N}{N}\right)^{m_++1}t^{(d-1)}\leq
\frac{e^{-m_+\gamma_+t_+}}{a_N}\frac{M_N}{N}e^{-\gamma_-t}t^{(d-1)}\leq
d_1e^{-\gamma_-t}t^{(d-1)}< 1/2.
$$
It is then
clear that (\ref{eq:trickbis}) holds for all $t$ in the region
(\ref{eq:region}).

The general case, where $\cal C$ is a finite set of points is
easily treated by noting that there exists $S\in\N_*$ so that
$S{\cal C}\subset \Z^{2d}$ so that
$$
d_{\R^{2d}}(y, {\cal C} +\Z^{2d})=S^{-1} d_{\R^{2d}}(Sy, S{\cal
C}+ S\Z^{2d})\geq S^{-1} d_{\R^{2d}}(Sy, \Z^{2d}).
$$
\endproof

\bigskip
\bigskip

{\bf Acknowledgments:} SDB whishes to thank S. Nonnenmacher for
helpful conversations and M. Belliart for guiding him to the
literature on ergodic toral automorphisms.

%%%%%%%%%%%%%%%%%%%%%%%%%%%%%%%%%%%%%%%%%%%%%%%%%%%%%%%%%%%%%%%%
%%%%%%%%%%%%%%%%%%%%%%%%%%%%%%%%%%%%%%%%%%%%%%%%%%%%%%%%%%%%%%%%
%%%%%%%%%%%%%%%%%%%%%%%%%%%%%%%%%%%%%%%%%%%%%%%%%%%%%%%%%%%%%%%%

\end{document}